\documentclass{IEEEtran}
\usepackage[T1]{fontenc}
\usepackage[latin9]{inputenc}
\usepackage{float}
\usepackage{amsmath}
\usepackage{amssymb}
\usepackage{graphicx}

\makeatletter

\floatstyle{ruled}
\newfloat{algorithm}{tbp}{loa}
\providecommand{\algorithmname}{Algorithm}
\floatname{algorithm}{\protect\algorithmname}

\usepackage{cite,balance,flushend}
\usepackage{tikz}
\usepackage{pgf}\usetikzlibrary{calc,arrows}
\usepackage{algorithm}
\usepackage{algpseudocode}

\makeatother

\begin{document}
\title{Reinforcement Learning based Per-antenna Discrete Power Control for
Massive MIMO Systems\thanks{The work was supported by the U.K. Engineering and Physical Sciences
Research Council (EPSRC) under Grants EP/P009549/1 and EP/P009670/1.}}
\author{Navneet Garg, Mathini Sellathurai$^{\dagger}$, Tharmalingam Ratnarajah\\
The University of Edinburgh, UK, $^{\dagger}$Heriot-Watt university,
Edinburgh, UK.}
\maketitle
\begin{abstract}
Power consumption is one of the major issues in massive MIMO (multiple input multiple output) systems, causing increased long-term operational cost and overheating issues. In this paper, we consider per-antenna power allocation with a given finite set of power levels towards maximizing the long-term energy efficiency of the multi-user systems, while satisfying the QoS (quality of service) constraints at the end users in terms of required SINRs (signal-to-interference-plus-noise ratio), which depends on channel information. Assuming channel states to vary as a Markov process, the constraint problem is modeled as an unconstraint problem, followed by the power allocation based on $Q$-learning algorithm. Simulation results are presented to demonstrate the successful minimization of power consumption while achieving the SINR threshold at users. 
\end{abstract}

\section{Introduction}

Massive MIMO systems are the central part of 5G and next generation
wireless networks. Due to large number of antennas in the array, the
increased power consumption i.e. reduced energy efficiency (EE), causes
increased operational cost and overheating problems which leads to
reduced lifespan of the array. The power allocation problem has been
widely investigated in literature via different schemes such as antenna
selection schemes \cite{7172496,7539316,7855725,8207624,8306705,8360626,8426033,8457308,8490388,8519787,8654616},
machine/deep learning (ML/DL) schemes \cite{8618345,8666158,8790780},
convex approximation based \cite{8744489,10.5555/2747012.2747124},
etc. In massive MIMO systems, transmit correlation with mutual coupling
is studied in \cite{6522419}, while with hybrid precoding, power
consumption cost is minimized in \cite{8295113}. The antenna selection
methods require NP-hard non-convex problem to be solved, and power
allocation step is still needed, which reduces its preference of usage
in practice. The drawback of ML/DL approaches is that they require
the huge data for training and the optimal solution is not guaranteed.
Convex-approximation based approaches approximate the non-convex EE
expressions into convex ones and obtain sub-optimal power allocation.
Therefore, a unified power allocation and antenna selection approach
is essential in improving the energy efficiency. 

In this paper, we present the discrete power allocation scheme using
reinforcement $Q$-learning for downlink multi-user massive MIMO system
towards that maximization of the long-term energy efficiency subject
to the total power constraint, per-antenna power constraint, and the
quality of service (QoS) constraints at the end users in terms of
SINR. Discrete power allocation can also be considered as a generalization
of antenna selection schemes, which has only two power levels. Assuming
the channel changes as a Markov process in the time-slotted model
with unknown transition probabilities, the long term energy efficiency
maximization problem is presented subjected to total power constraint
and per-antenna power constraint. This constraint problem is formulated
as an unconstraint problem and $Q$-learning is used to obtain the
solution. Simulation results demonstrate that $Q$-learning algorithm
converges and minimizes the power consumption, while satisfying the
QoS constraint at users.

\section{System Model}

Consider a downlink multi-user system, where a base station (BS) is
equipped with a large number of antennas ($M$). 
\begin{figure}
\centering\includegraphics[width=1\columnwidth]{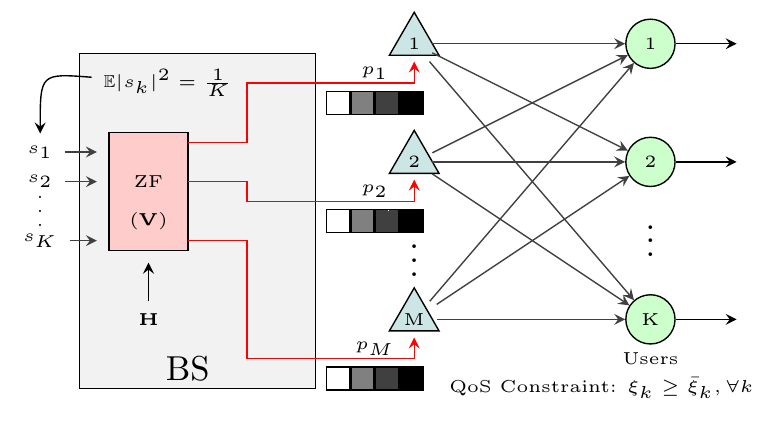}

\caption{BS with discrete power control $\mathbf{P}=\mathcal{D}\left\{ p_{1},\ldots,p_{M}\right\} $
and $\mathbb{E}\left\{ \mathbf{s}\mathbf{s}^{\dagger}\right\} =\frac{1}{K}\mathbf{I}_{K}$
and $\mathbb{E}\left\{ \mathbf{s}\right\} =\mathbf{0}$.\label{fig:BS-with-discrete}}
\end{figure}
The BS serves simultaneously a set of $K$ users indexed by $\mathcal{K}=\left\{ 1,\ldots,K\right\} $.
The transmitted signal from the BS can be expressed as 
\begin{equation}
\mathbf{x}=\mathbf{P}^{1/2}\sum_{k\in\mathcal{K}}\mathbf{v}_{k}s_{k}=\mathbf{P}^{1/2}\mathbf{V}\mathbf{s},
\end{equation}
where $\mathbf{s}=\left[s_{1},\ldots,s_{K}\right]^{T}$ is $K\times1$
symbol vector to be transmitted such that for each $k^{th}$ user,
$\mathbb{E}\left\{ s_{k}\right\} =0$, $\mathbb{E}\left\{ s_{k}s_{j}^{*}\right\} =\frac{1}{K}\delta_{kj}$
and $\mathbb{E}\left\{ \mathbf{s}\mathbf{s}^{\dagger}\right\} =\frac{1}{K}\mathbf{I}$
with $\delta_{kj}$ being the Kronecker delta having value 1 when
$k=j$ and $0$ otherwise; the matrix $\mathbf{V}=\left[\mathbf{v}_{1},\ldots,\mathbf{v}_{K}\right]$
is an $M\times K$ orthonormal precoder such that $\mathbf{V}^{\dagger}\mathbf{V}=\mathbf{I}_{K}$;
the quantity $\mathbf{P}=\mathcal{D}\left(p_{1},\ldots,p_{M}\right)$
is an $M\times M$ diagonal power allocation matrix with non-negative
entries. Using the above, the per-antenna and the total power constraints
at the BS can be obtained as 
\begin{align}
T_{per}(p_{m}) & =\left[\mathbb{E}\left\{ \mathbf{x}\mathbf{x}^{H}\right\} \right]_{m,m}\\
 & =\frac{p_{m}}{K}\left[\mathbf{V}\mathbf{V}^{H}\right]_{m,m}\leq\bar{P}_{m},\forall m\\
T_{tot}(\mathbf{P}) & =\mathbb{E}\left\Vert \mathbf{x}\right\Vert ^{2}=\frac{1}{K}\text{tr}(\mathbf{P}\mathbf{V}\mathbf{V}^{H})\leq\bar{P}_{T},\label{eq:tot_pow}
\end{align}
where $\bar{P}_{m}$ and $\bar{P}_{T}$ are the $m^{th}$ antenna
and the total power constraints. For simplicity, we assume equal power
constraint per antenna i.e. $\bar{P}_{m}=\bar{P}_{per},\forall m=1,\ldots,M$.
Towards discrete power control, let the set $\mathcal{P}=\left\{ p^{(1)},\ldots,p^{(\left|\mathcal{P}\right|)}\right\} $
denote all the power levels for each antenna i.e. $p_{m}\in\mathcal{P},\forall m$
and $\mathbf{P}\in\mathcal{P}^{M}$ such that $0=p^{(1)}\leq\cdots\leq p^{(\left|\mathcal{P}\right|)}=\bar{P}_{per}$,
where $\bar{P}_{per}$ also denotes the maximum power transmitted
by a single antenna. Let $\mathbf{h}_{k}$ denote the channel state
information (CSI) from BS at origin to the $k^{th}$ user. Through
this channel, the received signal at the $k^{th}$ user can be written
as 
\begin{align}
y_{k} & =\mathbf{h}_{k}^{H}\mathbf{x}+n_{k},\\
 & =\mathbf{h}_{k}^{H}\mathbf{P}^{1/2}\mathbf{V}\mathbf{s}+n_{k},\\
 & =\mathbf{h}_{k}^{H}\mathbf{P}^{1/2}\mathbf{v}_{k}s_{k}+\mathbf{h}_{k}^{H}\mathbf{P}^{1/2}\mathbf{V}_{-k}\mathbf{s}_{-k}+n_{k},
\end{align}
where $n_{k}\sim\mathcal{CN}(0,\sigma^{2})$ is the circularly symmetric
complex Gaussian noise; $\mathbf{V}_{-k}=\left[\mathbf{v}_{1},\ldots,\mathbf{v}_{k-1},\mathbf{v}_{k+1},\ldots,\mathbf{v}_{K}\right]$
and $\mathbf{s}_{-k}=\left[s_{1},\ldots,s_{k-1},s_{k+1},\ldots,s_{K}\right]^{T}$.
At the $k^{th}$user, the resultant SINR can be given as 
\begin{equation}
\xi_{k}(\mathbf{P}|\mathbf{H})=\frac{\left|\mathbf{h}_{k}^{H}\mathbf{P}^{1/2}\mathbf{v}_{k}\right|^{2}\frac{1}{K}}{\text{tr}\left(\mathbf{h}_{k}^{H}\mathbf{P}^{1/2}\mathbf{V}_{-k}\mathbf{V}_{-k}^{H}\mathbf{P}^{1/2}\mathbf{h}_{k}\right)\frac{1}{K}+\sigma^{2}},
\end{equation}
which depends on CSI $\mathbf{H}=\left[\mathbf{h}_{1},\ldots,\mathbf{h}_{K}\right]$.
Thus, stacking all the received signals gives $\mathbf{y}=\mathbf{H}^{H}\mathbf{P}^{1/2}\mathbf{V}\mathbf{s}+\mathbf{n}$.
If the CSI variations follow a Markov process, the resultant SINR
process will also be Markov. In other words, the power in the elements
of $\mathbf{P}$ needs to be adjusted according to CSI to satisfy
QoS constraints at the $k^{th}$ user. Further, the achievable sum-rate
is given as 
\begin{equation}
R(\mathbf{P}|\mathbf{H})=\sum_{k\in\mathcal{K}}\log_{2}\left(1+\xi_{k}(\mathbf{P}|\mathbf{H})\right).
\end{equation}
The resultant energy efficiency can be defined as the ratio of the
sum rate over the total power incurred in the transmission as 
\begin{equation}
\eta(\mathbf{P}|\mathbf{H})=\frac{R(\mathbf{P}|\mathbf{H})}{T_{tot}(\mathbf{P})},
\end{equation}
where the circuit power is ignored as it is a constant. In the following,
we simplify the sum rate for two popular precoding schemes based on
ZF (zero-forcing) and MRT (maximal ratio transmission).

\subsection{Zero-forcing}

For zero forcing transmission, to find the precoder satisfying $\mathbf{h}_{k}^{H}\mathbf{P}^{1/2}\mathbf{v}_{j}=0,\forall k\neq j$,
we normalize the columns of $\mathbf{V}'=\mathbf{P}^{1/2}\mathbf{H}\left(\mathbf{H}^{H}\mathbf{P}\mathbf{H}\right)^{-1}$
to be unit norm columns. The above precoder results in the received
signal $y_{k}=\mathbf{h}_{k}^{H}\mathbf{P}^{1/2}\mathbf{v}_{k}s_{k}+n_{k}$,
resulting into the sum rate

\begin{equation}
R_{ZF}(\mathbf{P}|\mathbf{H})=\sum_{k\in\mathcal{K}}\log_{2}\left(1+\frac{\left|\mathbf{h}_{k}^{H}\mathbf{P}^{1/2}\mathbf{v}_{k}\right|^{2}}{\sigma^{2}K}\right).
\end{equation}

\subsection{Maximal ratio transmission}

For MRT based precoding, the precoder is set as $\mathbf{v}_{k}=\frac{\mathbf{P}^{1/2}\mathbf{h}_{k}}{\sqrt{\mathbf{h}_{k}^{H}\mathbf{P}\mathbf{h}_{k}}}$.
Note that MRT precoding is used for low complexity operations, thus,
the precoding vectors are not orthonormalized. The sum rate can be
simplified as 
\begin{align*}
 & R_{MRT}(\mathbf{P}|\mathbf{H})=\sum_{k\in\mathcal{K}}\log_{2}\left(1+\frac{\mathbf{h}_{k}^{H}\mathbf{P}\mathbf{h}_{k}}{\sum_{j\neq k}\frac{\mathbf{h}_{k}^{H}\mathbf{P}\mathbf{h}_{j}\mathbf{h}_{j}^{H}\mathbf{P}\mathbf{h}_{k}}{\mathbf{h}_{j}^{H}\mathbf{P}\mathbf{h}_{j}}+K\sigma^{2}}\right).
\end{align*}

\subsection{Problem formulation}

Our goal is to maximize the energy efficiency of transmissions via
discrete power allocations. However, note that in each time slot,
finding discrete power levels for each antenna in the massive MIMO
system is an NP-hard search problem and a non-convex problem. Moreover,
the estimation of CSI in massive MIMO consumes resources. Therefore,
for faster operations, utilizing the CSI correlation via Markov process,
reinforcement learning is utilized to obtain these power levels. Thus,
assuming the channel information varies as a finite state Markov chain,
our objective to find the discrete power allocation to maximize the
long term efficiency subject to the QoS constraints satisfied for
each user, can be expressed as 
\begin{align}
\max_{\mathbf{P}(t)} & \sum_{\tau=t}^{\infty}\gamma^{\tau-t}\eta(\mathbf{P}(t)|\mathbf{H}(t))\\
\text{subject to } & T_{tot}(\mathbf{P}(t),\mathbf{H}(t))\leq\bar{P}_{T},\mathbf{P}(t)\in\mathcal{P}^{M},\nonumber \\
 & \xi_{k}(\mathbf{P}(t)|\mathbf{H}(t))\geq\bar{\xi}_{k},\forall k\in\mathcal{K},
\end{align}
where $\bar{\xi}_{k}$ represents the SINR requirements for QoS at
$k^{th}$ user, and $(t)$ denotes their time dependent behavior.
Note that the total power is also considered here a function of $\mathbf{H}$.
It is due to the fact that the precoders are computed using channel
information $\mathbf{H}$. This makes the problem non-convex and difficult
to solve.

\section{Reinforcement Learning}

\subsection{Dynamics of EEPA}

We consider time varying channel across time slots. Within a time
slot, the channel remains constant. The CSI in a cellular network
varies if the user is walking, running or in a vehicle. In literature
\cite{8626185,8746268}, the time varying channel is modeled using
a finite state Markov chain, where the ergodic channel in each time
slot takes value in one of the Markov states. Let $\mathcal{H}=\left\{ \mathbf{H}^{(1)},\ldots,\mathbf{H}^{(\left|\mathcal{H}\right|)}\right\} $
denote the states in the Markov chain. The transition probability
between channel states is fixed and unknown\footnote{In some literature, first order auto-regressive process is used to
model the channel variations due to the mobility, where the resulting
channel model provides continuous state Markov process, rather than
finite state chain.}. 

\subsection{States, Actions and Rewards}

For the above system dynamics, let $\underline{s}(t)$ be the state
at time $t$, which is given as the CSI of the same slot as $\underline{s}(t)=\mathbf{H}(t)\in\mathcal{H}$.
An action in the system corresponds to discrete power control i.e.
$\underline{a}(t)=\mathbf{P}(t)\in\mathcal{P}^{M}$. The action chosen
is evaluated using the reward which is defined as the energy efficiency
i.e. 
\begin{equation}
\underline{r}(\underline{s}(t),\underline{a}(t))=\frac{1}{\left(\sum_{k\in\mathcal{K}}\left|\xi_{k}(\mathbf{P}(t)|\mathbf{H}(t))-\bar{\xi}_{k}\right|\right)T_{tot}(\mathbf{P}(t),\mathbf{H}(t))},
\end{equation}
where $\left|\cdot\right|$ ensures that the resulting SINR does not
achieve values far from $\bar{\xi}_{k}$.

Here, the learner seeks the optimum action $\underline{a}(t)$ based
on the previous observation $\mathbf{H}(t-1)=\underline{s}(t-1)$
by interactively making sequential decisions and observing the corresponding
costs. In this way, the agent learns the best action policy against
the random Markov chain transitions. Let the policy function be $\pi:\mathcal{H}\rightarrow\mathcal{P}^{M}$,
which maps a state to an action. Under policy $\pi(\cdot)$, the power
allocation is carried out via action $\underline{a}(t+1)=\pi(\underline{s}(t))$,
dictating the allocation policy at time $t+1$. For the reward $\underline{r}_{\pi}\left(\underline{s}(t)\right)=\underline{r}\left(\underline{s}(t),\pi\left(\underline{s}(t)\right)\right)$,
power consumption performance is measured through the state value
function as 
\begin{equation}
V_{\pi}(\underline{s}(t))=\sum_{\tau=t}^{\infty}\gamma^{\tau-T}\underline{r}_{\pi}(\underline{s}(t)),
\end{equation}
which is the total average cost incurred over an infinite time horizon.
The objective of this paper is to find the optimal policy $\pi^{*}$
such that the average cost of any state is maximized 
\[
\pi^{*}=\arg\max_{\pi}V_{\pi}(\mathbf{S}).
\]

\subsection{Action set reduction}

For the BS equipped with $M$ antennas, there are huge number of $\left|\mathcal{P}\right|^{M}$
possible actions. However, not all actions are valid actions. Valid
actions are those actions which satisfy the power constraint in \eqref{eq:tot_pow}.
The total power $T_{tot}(\mathbf{P},\mathbf{H})$ depends on the normalized
precoder $\mathbf{V}$. To simplify the constraint in order to reduce
the valid action set, we approximate the total power constraint as
\begin{equation}
\frac{1}{K}\text{tr}(\mathbf{P}\mathbf{V}\mathbf{V}^{H})\approx\frac{1}{K}\text{tr}\left(\mathbf{P}\mathbb{E}\left\{ \mathbf{V}_{R}\mathbf{V}_{R}^{H}\right\} \right)=\frac{1}{M}\text{tr}\left(\mathbf{P}\right)\leq\bar{P}_{T},\label{eq:pow_con_red}
\end{equation}
where $\mathbf{V}_{R}$ is any random orthonormal precoder; the equality
on the right follows from \cite[Lem. 1]{8379356}. Further, at least
$K$ actions should be non-zero i.e. $p_{i_{k}}>0,\forall k\in\mathcal{K}$
that excludes $\sum_{k=1}^{K-1}{M \choose k}$ actions in $\mathcal{P}^{M}$.
To get the minimum transmission power constraint to reduce huge number
of possibilities, we approximate the QoS constraint as 
\begin{align*}
\xi_{k}(\mathbf{P}|\mathbf{H}) & =\frac{\left|\mathbf{h}_{k}^{H}\mathbf{P}^{1/2}\mathbf{v}_{k}\right|^{2}}{\text{tr}\left(\mathbf{h}_{k}^{H}\mathbf{P}^{1/2}\mathbf{V}_{-k}\mathbf{V}_{-k}^{H}\mathbf{P}^{1/2}\mathbf{h}_{k}\right)+K\sigma^{2}},\\
 & \stackrel{(a)}{\approx}\frac{\text{tr}\left(\mathbf{P}\mathbf{v}_{k}\mathbf{v}_{k}^{H}\right)}{\text{tr}\left(\mathbf{P}\mathbf{V}_{-k}\mathbf{V}_{-k}^{H}\right)+K\sigma^{2}},\\
 & \stackrel{(b)}{\approx}\frac{\text{tr}\left(\mathbf{P}\right)\frac{1}{M}}{\text{tr}\left(\mathbf{P}\right)\frac{K-1}{M}+K\sigma^{2}}=\frac{1}{(K-1)+KM\frac{\sigma^{2}}{\text{tr}\left(\mathbf{P}\right)}},
\end{align*}
where (a) follows from the massive MIMO channel hardening effect $\mathbf{h}_{k}\mathbf{h}_{k}^{H}\rightarrow\mathbf{I}_{M}$;
and (b) follows similarly from \eqref{eq:pow_con_red}. For ZF precoding,
we have $\frac{1}{KM\frac{\sigma^{2}}{\text{tr}\left(\mathbf{P}\right)}}\geq\bar{\xi}_{k}$
$\implies$ $\text{tr}\left(\mathbf{P}\right)\geq KM\sigma^{2}\bar{\xi}_{k}$.
Let $\bar{P}_{\min}$ denote this lower bound on the transmission
power. The new action space can now be expressed as 
\begin{equation}
\bar{\mathcal{P}}_{M}=\left\{ \left(\begin{array}{c}
p_{1}\\
\vdots\\
p_{M}
\end{array}\right):\begin{array}{c}
\bar{P}_{\min}\leq\text{tr}(\mathbf{P}(t))\leq M\bar{P}_{T},\\
p_{i_{k}}>0,\forall k\in\mathcal{K}
\end{array}\right\} ,
\end{equation}
where $\bar{\mathcal{P}}_{M}\subset\mathcal{P}^{M}$. Note that the
above approximations are to reduce the possible actions, and thus,
it does not affect the optimal power allocations. 

\subsection{Bellman's Equations and $Q$-learning}

Let $\Pr(\underline{s},\underline{s}'|\underline{a})$ be the probability
of transition from the current state $\underline{s}$ to the next
state $\underline{s}'$ under action $\underline{a}$. Bellman equations
express the state value functions in a recursive fashion as 
\begin{align}
V_{\pi}(\underline{s})= & \underline{r}_{\pi}(\underline{s})+\gamma\sum_{\underline{s}'\in\Xi^{K}}\Pr(\underline{s},\underline{s}'|\pi(\underline{s}))V_{\pi}(\underline{s}'),\forall\underline{s}\\
Q_{\pi}(\underline{s},\underline{a})= & \underline{r}_{\pi}(\underline{s})+\gamma\sum_{\underline{s}'\in\Xi^{K}}\Pr(\underline{s},\underline{s}'|\underline{a})V_{\pi}(\underline{s}'),\forall\underline{s},\underline{a}.\label{eq:Q-fun1}
\end{align}
The above equations can be used to obtain the optimal policy by minimizing
$Q$-function as 
\begin{equation}
\pi^{*}=\arg\max_{\underline{a}}Q_{\pi}(\underline{s},\underline{a}),\forall\underline{s},
\end{equation}
where under $\pi^{*}$, $V_{\pi^{*}}(\underline{s})=\max_{\underline{a}}Q_{\pi}(\underline{s},\underline{a})$
and it gives the solution 
\begin{align*}
Q_{\pi}(\underline{s},\underline{a}) & =\underline{r}_{\pi}(\underline{s})+\gamma\sum_{\underline{s}'\in\Xi^{K}}\Pr(\underline{s},\underline{s}'|\underline{a})\max_{\underline{a}}Q_{\pi}(\underline{s},\underline{a}),\\
 & =\sum_{\underline{s}'\in\Xi^{K}}\Pr(\underline{s},\underline{s}'|\underline{a})\left[\underline{r}(\underline{s},\underline{a})+\gamma\max_{\underline{a}}Q_{\pi}(\underline{s},\underline{a})\right].
\end{align*}
The above solution demands an iterative solution for $Q$-function,
which is given in Algorithm \ref{alg:Hit-probability-Maximization}.
\begin{algorithm}[t]
\begin{algorithmic}[1]

\Require{ state $\underline{s}(0)$ randomly and $Q_{0}(\underline{s},\underline{a})=0\forall\underline{s},\underline{a}$
}

\For{ $t=1,2,\ldots$,  }

\State{For given profile $\underline{s}(t-1)$, take action $\underline{a}(t)$
as 
\[
\underline{a}(t)=\begin{cases}
\arg\max_{\underline{a}}Q_{t-1}(\underline{s}(t),\underline{a}) & \text{w.p.}\:1-\epsilon\\
\text{random }\underline{a}\in\bar{\mathcal{P}} & \text{w.p.}\:\epsilon
\end{cases}
\]
}

\State{Observe $\underline{s}(t)$ and compute $\underline{r}(\underline{s}(t),\underline{a}(t))$}

\State{Update 
\begin{align}
 & Q_{t}(\underline{s}(t),\underline{a}(t))=(1-\beta_{t})Q_{t-1}(\underline{s}(t),\underline{a}(t))\\
 & +\beta_{t}\left[\underline{r}(\underline{s}(t),\underline{a}(t))+\gamma\max_{\underline{a}}Q_{t-1}(\underline{s}(t),\underline{a})\right].\nonumber 
\end{align}
}

\EndFor

\end{algorithmic}

\caption{$Q$-learning algorithm.\label{alg:Hit-probability-Maximization}}
\end{algorithm}

In a time slot $t$, after observing the state $\underline{s}(t)$,
the $\epsilon$-greedy action $\underline{a}(t)$ is taken and instantaneous
cost $\underline{r}(\underline{s}(t),\underline{a}(t))+\gamma\max_{\underline{a}}Q(\underline{s}(t),\underline{a})$
is incurred. Under mean squared error (MSE) criteria, the MSE expression
for the estimated $Q$-function values can be written as 
\begin{align*}
\epsilon(\underline{s}(t),\underline{a}(t)) & =\Bigg[\underline{r}(\underline{s}(t),\underline{a}(t))+\\
 & \hfill\gamma\max_{\underline{a}}Q(\underline{s}(t),\underline{a})-Q(\underline{s}(t),\underline{a}(t))\Bigg]^{2}.
\end{align*}
Minimizing the above error expression for $Q$-values using gradient
descent method yields the following 
\begin{align}
 & Q_{t}(\underline{s}(t),\underline{a}(t))=(1-\beta_{t})Q_{t-1}(\underline{s}(t),\underline{a}(t))\\
 & +\beta_{t}\left[\underline{r}(\underline{s}(t),\underline{a}(t))+\gamma\max_{\underline{a}}Q_{t-1}(\underline{s}(t),\underline{a})\right],\nonumber 
\end{align}
where $Q_{t}$ is estimated $Q$-values at time $t$. It can be noted
that the convergence of the algorithm depends on the values of $\beta_{t}$.
Choosing $\beta_{t}$ such that $\sum_{t}\beta_{t}<\infty$ guarantees
the convergence. These cases of convergence and several other related
algorithms has been thoroughly studied in \cite{Tsitsiklis1997}. 

Note that the cardinality of action space is increased exponentially
for increase in the number of antennas and the number of power levels.
Therefore, to make it scalable, deep reinforcement learning based
methods will be investigated as a part of future work. 

\section{Simulation Results}

The following values are assumed for $Q$-learning parameters: $M=8,16$
antennas; $K=4$ downlink users $\left|\mathcal{P}\right|=3,5$; $1000$
number of episodes for $Q$-learning with each episode having $2000$
iterations; exploration decay factor per episode $0.1$; transmit
power constraint $28$ dB; per-antenna maximum power constraint $30$
dB; $Q$oS constraint for SINR $20$ dB; number of channel states
$\left|\mathcal{H}\right|=128$. Zero-forcing based precoding is assumed
since $M$ is not high enough and the present $Q$-learning algorithm
is computationally time consuming. 

\begin{figure}
\centering\includegraphics[width=1\columnwidth]{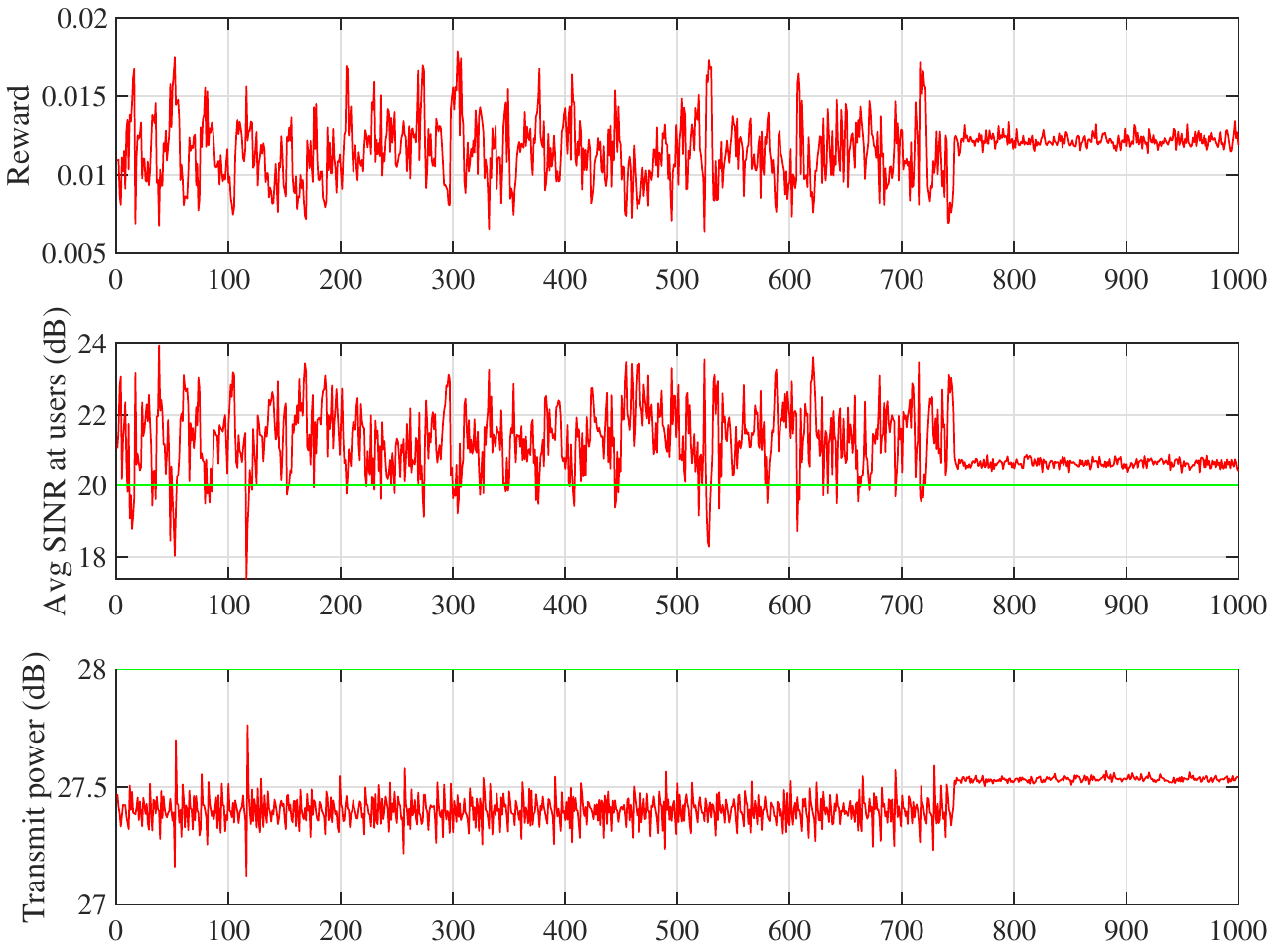}

\caption{Progress of average rewards, average SINR at users, and average transmit
power at BS for different iterations for $M=16$ and $\left|\mathcal{P}\right|=3$
levels with per-antenna constraint, total transmit power constraint
and user-SINR constraint of $30$ dB, $28$ dB and $20$ dB (green
lines) respectively. \label{fig:Progress-of-average}}
\end{figure}
\begin{figure}
\centering\includegraphics[width=1\columnwidth]{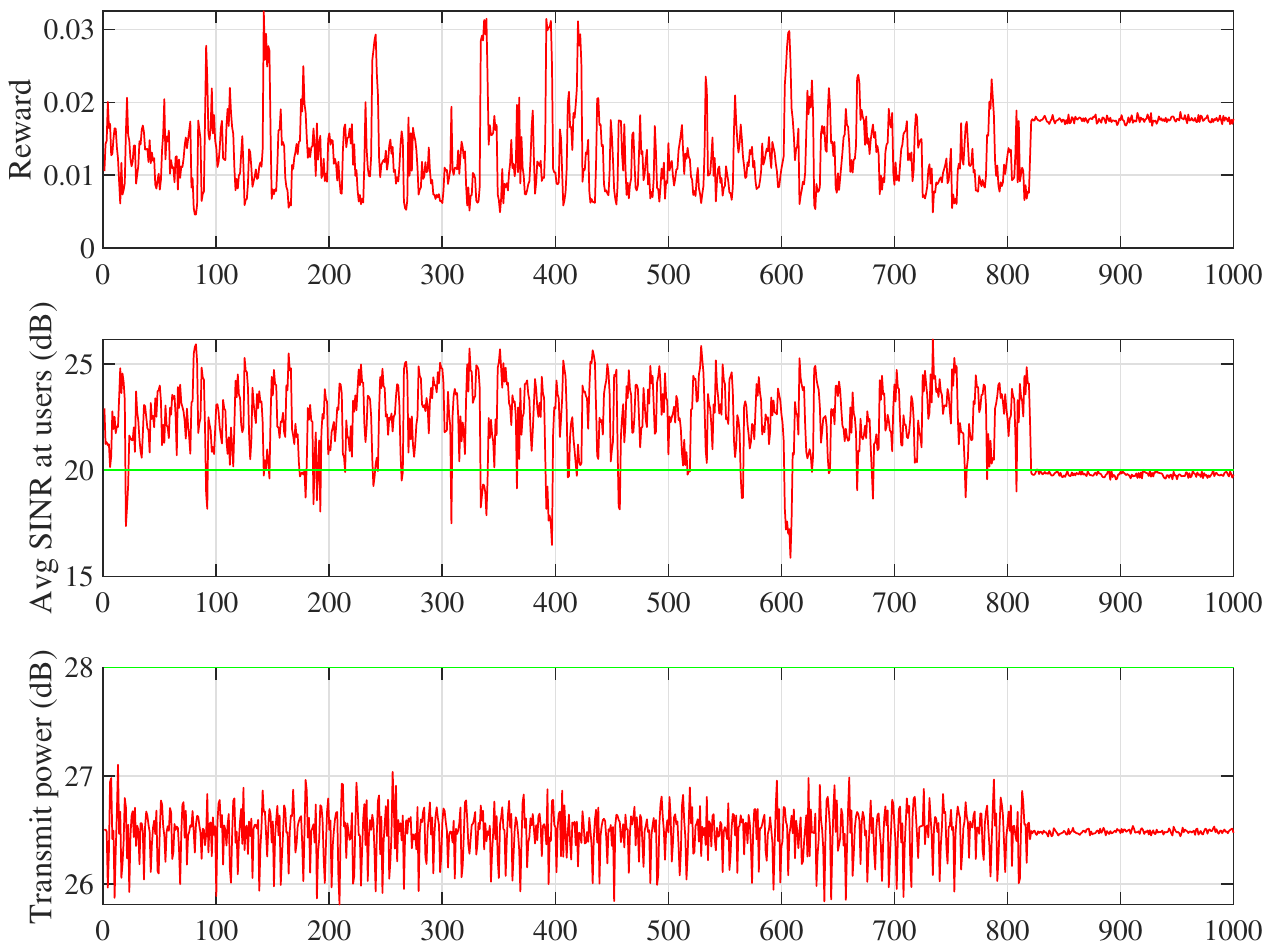}

\caption{Progress of average rewards, average SINR at users, and average transmit
power at BS for different iterations for $M=8$ and $\left|\mathcal{P}\right|=5$
levels with per-antenna constraint, total transmit power constraint
and user-SINR constraint of $30$ dB, $28$ dB and $20$ dB respectively.
. \label{fig:Progress-of-average-1}}
\end{figure}

Figure \ref{fig:Progress-of-average} shows the plots for the progresses
of average reward over iterations, average SINR across users and iterations,
and average transmit power across iterations, respectively for $M=16$
antennas at BS and $\left|\mathcal{P}\right|=3$ power levels for
each antenna. The action set is reduced from $3^{16}$ to around $12000$
entries. It can be seen that the $Q$-learning learns the optimum
power allocation in terms of reward, and the learned actions provide
SINR greater than the $Q$oS constraint for each user, keeping the
transmit power within the constraint. Due to larger size of $Q$-matrix,
it takes around 750 iterations to learn the optimum converging action.
Similar trends can be seen for the case, when five power levels are
assumed as shown in Figure \ref{fig:Progress-of-average-1}. It shows
the successful application of $Q$-learning in quickly finding the
optimum power allocation among such a large set of possibilities $3^{16}\approx4\times10^{7}$. 

\section{Conclusion}

In this paper, we have presented reinforcement learning solution for
discrete power allocation, which is a combinatorial optimization problem
and is NP-hard. By leveraging the correlation between channels for
slowing moving scenarios in wireless cellular networks, we model the
channel variations as a finite state Markov chain and presented the
RL formulation where the constraints are transmit power constraint
and the Quality of service guarantee in terms of received SINR at
each user with an objective of maximizing the energy efficiency at
the transmitter. Typically, to handle the constraints in $Q$-learning,
primal-dual approaches are used. However, we model the reward function
to incorporate these constraints, without needing any additional dual
variables in design. Simulations shows the successful application
of the power allocation while satisfying these constraints. 

The future work is to make the algorithm scalable for larger number
of power levels and larger number of antennas. 

\bibliographystyle{IEEEtran}
\bibliography{pa1}

\end{document}